# Reconfigurable Intelligent Surfaces: Architectures, Models, and Applications

Mojtaba Fallahzadeh, Mohammad Hossein Koohi Ghamsari, Alireza Mallahzadeh, Amirhossein Ghasemi, Gabriele Gradoni, Rahim Tafazolli

5GIC & 6GIC, Institute for Communication Systems (ICS), University of Surrey, Guildford, United Kingdom

*Abstract*— **Reconfigurable Intelligent Surfaces (RIS), also known as programmable metasurfaces, have emerged as a promising technology for next-generation wireless communication systems by enabling programmable control over the propagation environment. Through dynamic manipulation of the phase, amplitude, and polarization of electromagnetic waves, RIS introduces new degrees of freedom for enhancing coverage, spectral efficiency, energy efficiency, and interference management. This article reviews recent advances in RIS-enabled wireless communications, focusing on RIS architectures, emerging design and analysis techniques, and their impact on system-level performance. Key challenges related to scalability, hardware non-idealities, channel modeling, and control overhead are discussed, along with promising solutions and future research directions toward practical RIS deployment in beyond-5G and 6G networks.**

## I. Introduction and Scope of the Review

Reconfigurable Intelligent Surfaces (RIS), also known as programmable metasurfaces, have recently emerged as a disruptive paradigm for future wireless communication systems. By enabling software-controlled manipulation of electromagnetic waves through spatially distributed meta-atoms, RIS transforms the wireless propagation environment from a passive medium into an active and reconfigurable entity [1], [2]. This capability introduces new degrees of freedom for enhancing coverage, improving spectral and energy efficiency, and mitigating interference, all of which are critical requirements for beyond-5G and 6G networks [4], [7]. As a result, RIS has rapidly gained attention as a key building block for smart radio environments and network-level performance optimization [5], [6].

A substantial body of literature has already investigated the integration of RIS into wireless communication systems, with particular emphasis on channel modeling, beamforming strategies, and signal processing techniques [6]–[8]. These works primarily adopt a communication-theoretic viewpoint, often relying on simplified or idealized RIS models to enable tractable analysis at the system level. While such models are valuable for understanding network-level performance trends, they typically abstract away the physical realization of RIS, including the underlying electromagnetic structure, unit-cell design constraints, and hardware-induced non-idealities [9], [10]. This abstraction gap becomes increasingly critical as RIS transitions from conceptual studies toward practical deployment scenarios.

In parallel, extensive electromagnetic research has focused on the design, analysis, and characterization of metasurface unit cells, addressing topics such as tunability mechanisms, dispersion behavior, radiation leakage, losses, and fabrication constraints [11]–[13]. However, these studies are often detached from system-level considerations and are presented in a manner that may be less accessible to the wireless communications community. Consequently, a disconnect persists between electromagnetic-level RIS realizations and their impact on communication system performance, which limits the effective co-design of RIS hardware and wireless algorithms [4], [7].

Motivated by this gap, the objective of this review is to provide a unified and system-oriented perspective on RIS that bridges electromagnetic design principles with wireless communication performance. Rather than offering an exhaustive survey of theoretical models or fabrication techniques, this article emphasizes RIS architectures, emerging design and analysis paradigms, and their implications for practical wireless systems. Special attention is given to how different RIS realizations influence key metrics such as reconfigurability, scalability, efficiency, and controllability, all of which are essential for real-world deployment [5], [6].

The scope of this review encompasses three main aspects. First, the article discusses how RIS enables performance improvements in wireless communication systems, highlighting representative use cases and gains reported in recent studies [3], [4]. Second, emerging design and analysis techniques for RIS architectures are examined, with an emphasis on practical implementation considerations rather than purely theoretical formulations [11]–[13]. Finally, fundamental challenges, open research problems, and promising future directions are outlined, including hardware limitations, modeling accuracy, integration with intelligent control, and standardization issues [8], [10]. Through this structured discussion, the review aims to provide researchers and practitioners with a coherent roadmap for understanding the current state of RIS technology and its role in future wireless networks.

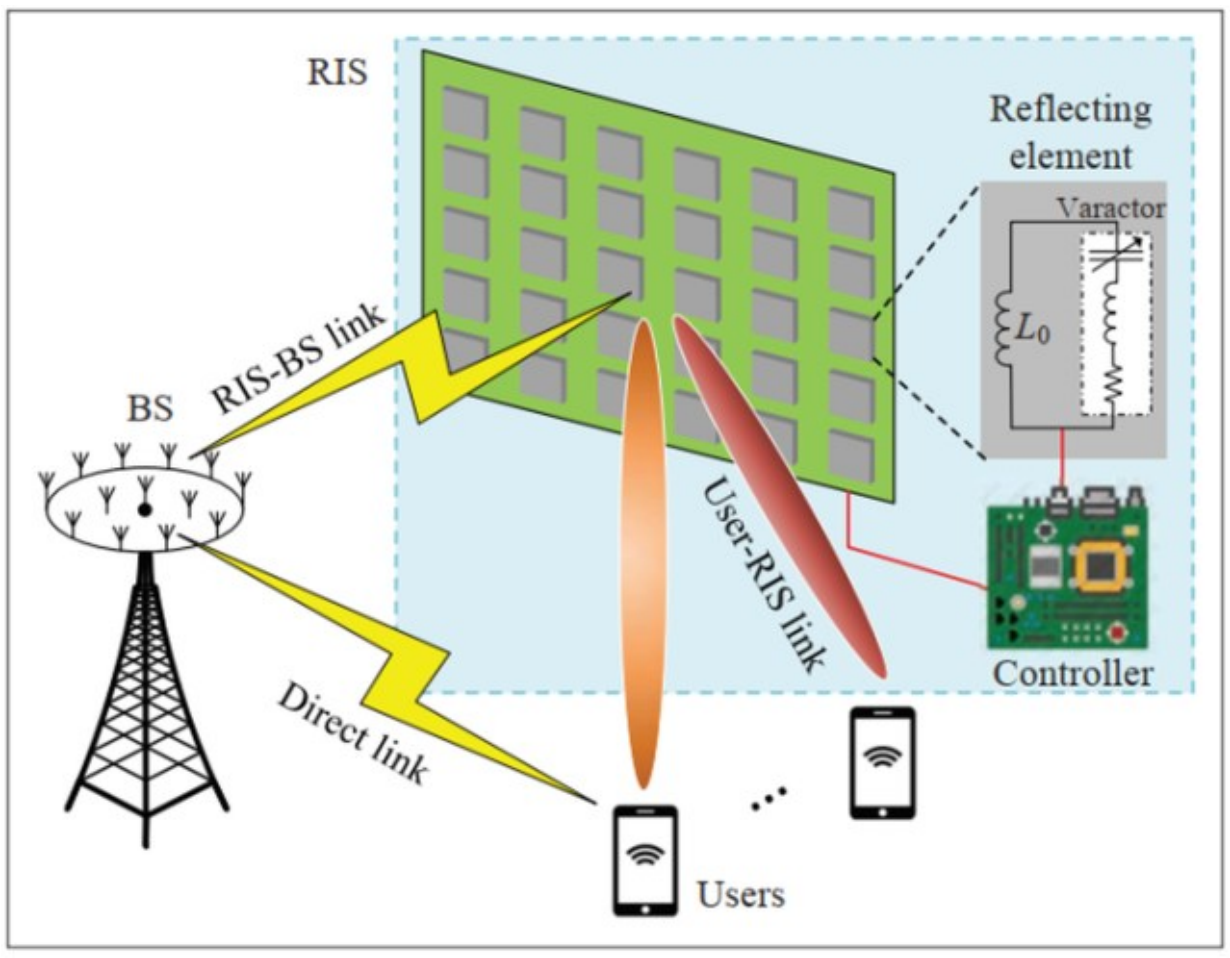

Fig. 1. A RIS-assisted massive MIMO system [5].

## II. Improvements in Wireless Communications With Programmable Metasurfaces

Reconfigurable Intelligent Surfaces (RIS) have demonstrated the ability to significantly enhance wireless communication performance by enabling programmable control over the radio propagation environment. Unlike conventional techniques that rely on increasing transmit power, deploying denser base stations, or employing more complex signal processing at the transceiver, RIS introduces a fundamentally different approach by reshaping the wireless channel itself. Through the coordinated adjustment of a large number of low-cost and mostly passive elements, RIS can tailor wave reflection, transmission, and scattering patterns to favor desired communication links while suppressing detrimental effects such as blockage and interference [1], [2].

One of the most widely reported benefits of RIS is coverage enhancement, particularly in scenarios characterized by severe blockage or non-line-of-sight (NLoS) propagation. By intelligently redirecting incident signals toward shadowed users, RIS effectively creates virtual line-of-sight paths that can substantially improve received signal power at the user side [3], [4]. Numerous experimental and simulation studies have reported signal-to-noise-ratio (SNR) gains on the order of several decibels, even when using low-resolution phase control at the RIS elements [6]. Such gains are especially valuable in densely built urban environments and indoor scenarios, where conventional solutions such as relays or additional access points may be impractical or costly.

In addition to coverage improvement, RIS plays a key role in enhancing spectral efficiency by enabling more favorable channel conditions for multi-user transmission. By shaping reflection patterns toward intended receivers, RIS can reduce multi-user interference and improve spatial selectivity, thereby increasing achievable data rates [7], [8]. When combined with advanced beamforming techniques at the base station, RIS can complement massive multiple-input multiple-output (MIMO) systems and extend their effective aperture without introducing additional active radio-frequency chains. This synergy between RIS and existing MIMO architectures is regarded as a promising pathway for scaling wireless networks toward the performance demands of beyond-5G and 6G systems [9].

Another major advantage of RIS is its potential to significantly improve energy efficiency. Since most RIS implementations rely on passive or semi-passive elements that consume negligible power compared to active relays, substantial performance gains can be achieved without a proportional increase in network energy consumption [4], [6]. This makes RIS particularly attractive for green communications and large-scale deployments, where operational cost and sustainability are critical design considerations. Studies have shown that even 1-bit or 2-bit phase-quantized RIS designs can capture most of the theoretically achievable performance gains, with only minor degradation compared to ideal continuous-phase configurations [7]. This observation motivates practical designs that favor simplicity, scalability, and low hardware complexity.

RIS has also been shown to mitigate various channel impairments inherent to dynamic wireless environments. By adaptively manipulating the reflected wavefront, RIS can reduce the severity of multipath fading, suppress Doppler effects in high-mobility scenarios, and enhance link reliability [8], [10]. These capabilities are particularly relevant for emerging applications such as vehicular communications, unmanned aerial vehicle (UAV) networks, and high-frequency millimeter-wave and terahertz systems, where channel variations are rapid and severe. In such contexts, RIS provides an additional layer of adaptability that complements traditional physical-layer and network-layer solutions [7].

From a system integration perspective, RIS enables the realization of smart radio environments, where the propagation medium becomes an integral part of the communication system design rather than an uncontrollable external factor [6]. This paradigm shift allows network operators to optimize performance metrics beyond conventional link adaptation, including secrecy capacity, localization accuracy, and latency [1], [9]. For instance, RIS can be configured to intentionally degrade signal reception at unintended locations, thereby enhancing physical-layer security without relying solely on cryptographic techniques [8]. Similarly, by controlling multipath characteristics, RIS can support more accurate positioning and sensing functionalities, which are expected to be tightly integrated with communication in future 6G networks [2], [7].

Overall, the improvements enabled by programmable metasurfaces span a wide range of wireless performance metrics, including received signal power, spectral efficiency, energy efficiency, reliability, and security. Importantly, many of these gains can be achieved using simple and scalable RIS architectures with limited phase resolution and straightforward control mechanisms [6], [7]. This highlights the strong potential of RIS as a practical and cost-effective technology for enhancing wireless communications and motivates continued research into architectures, design methodologies, and control strategies that jointly consider electromagnetic realizations and system-level performance [1], [9].

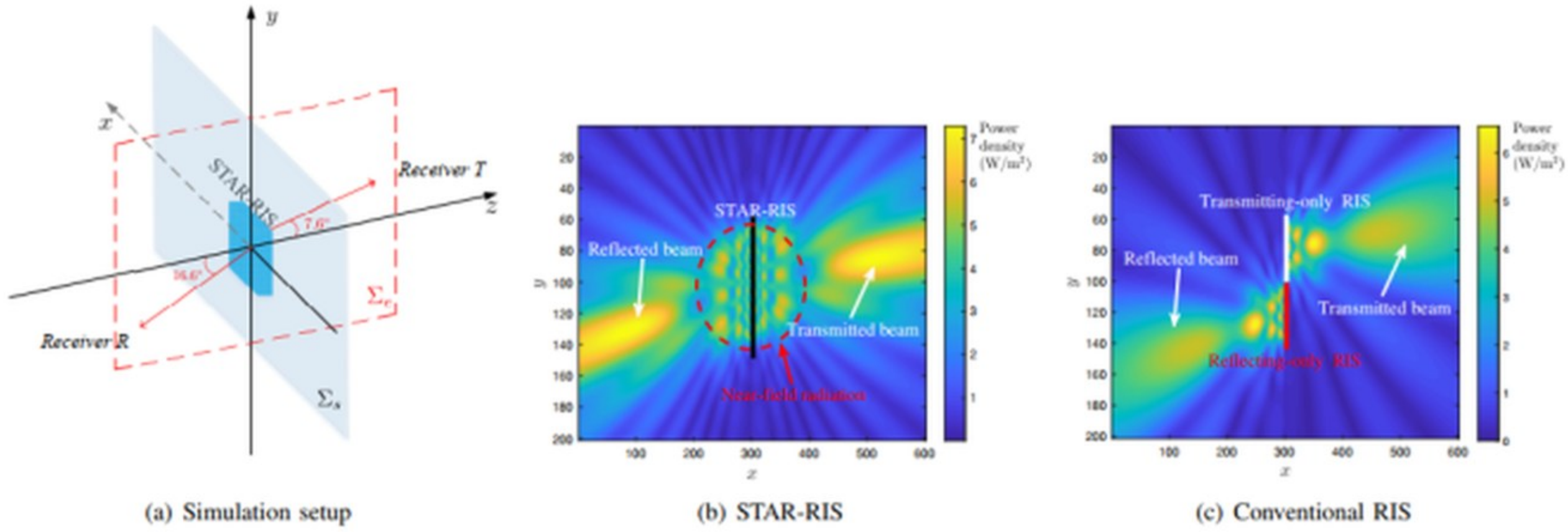

Fig. 2. Illustration of the simulation setup and simulated radiation coverage observed [3].

## III. Emerging Design and Analysis Techniques

The rapid development of reconfigurable intelligent surfaces has driven the emergence of new design and analysis techniques that extend beyond conventional antenna and propagation modeling frameworks. Unlike traditional fixed electromagnetic structures, RIS must simultaneously satisfy hardware-level electromagnetic constraints and system-level wireless performance objectives. This dual requirement has motivated a shift toward hybrid methodologies that combine physics-based electromagnetic modeling with communication-centric abstractions, enabling scalable analysis and practical optimization for large-scale deployments [1], [2].

From a design perspective, RIS architectures are commonly approached through two complementary paradigms: physics-based (electromagnetic) design and communication-based (system-level) design. In the physics-based paradigm, RIS is treated as a metasurface device whose unit cells are engineered to realize specific electromagnetic transformations, such as anomalous reflection, refraction, beam focusing, or polarization control [3], [4]. This approach emphasizes parameters such as unit-cell geometry, tunability mechanisms, impedance bandwidth, and radiation efficiency, typically evaluated through full-wave simulations or circuit-based equivalent models. While highly accurate, purely physics-based optimization becomes computationally prohibitive as the surface size grows and the degrees of freedom increase.

In contrast, communication-based design abstracts the RIS as a controllable surface characterized by phase shifts, reflection coefficients, or scattering responses, which can be readily integrated into wireless system models [5]–[7]. Such abstractions enable tractable optimization of beamforming, resource allocation, and network-level performance metrics but often rely on idealized assumptions that ignore non-ideal hardware effects, losses, and mutual coupling. As RIS moves toward practical realization, the limitations of overly simplified models become increasingly apparent, especially in scenarios involving wideband operation, oblique incidence, and near-field interactions [8], [9].

To balance accuracy and scalability, emerging analysis techniques increasingly incorporate ray-optics-inspired and wave-optics-based formulations as complementary tools. Ray-optics models offer intuitive and computationally efficient representations of wave redirection and are particularly attractive for large surfaces operating in electrically large environments [6], [7]. However, they fail to capture essential wave phenomena such as diffraction, coupling between elements, and resonant effects that are intrinsic to metasurface behavior. Wave-optics approaches, including full-wave solvers and modal analysis, provide higher fidelity but at a significantly increased computational cost [10]. Hybrid strategies that combine these approaches allow designers to capture critical electromagnetic effects while maintaining tractable simulation complexity.

Another important trend in RIS design and analysis is the growing adoption of data-driven and learning-based techniques. Machine learning methods, particularly deep neural networks, have been employed to accelerate unit-cell design, surrogate modeling, and inverse optimization of metasurface geometries [11]–[13]. By learning the complex mapping between physical design parameters and electromagnetic responses, these models significantly reduce reliance on repeated full-wave simulations while preserving reasonable accuracy.

At the system level, learning-assisted optimization has also been explored to address challenges such as RIS configuration, channel estimation, and joint transceiver–RIS optimization [5], [8]. In particular, adaptive and low-complexity data-driven schemes are attractive for dynamic environments, where channel conditions and user locations vary rapidly. These techniques enable adaptive control of RIS configurations with reduced overhead, paving the way for practical real-time operation in dense and mobile wireless networks [7].

Overall, emerging RIS design and analysis techniques highlight a clear trend toward co-design methodologies that tightly couple electromagnetic realizations with wireless system objectives. Rather than treating RIS as either a purely electromagnetic device or an abstract communication element, modern frameworks seek to unify these viewpoints through multilevel modeling, hybrid simulation strategies, and learning-assisted optimization [1], [6]. Such approaches are essential for bridging the gap between theoretical performance gains and deployable RIS-enabled wireless systems, and they play a crucial role in shaping the design philosophy of future smart radio environments [2], [9].

## IV. Challenges, Promising Solutions, and Future Trends

Despite the significant performance gains enabled by reconfigurable intelligent surfaces, several fundamental challenges must be addressed before RIS can be widely adopted in practical wireless networks. These challenges span multiple layers of the system, ranging from hardware scalability and electromagnetic non-idealities to channel estimation, control overhead, and real-time optimization. Understanding these limitations is essential for assessing the true deployment potential of RIS-enabled smart radio environments [1], [2].

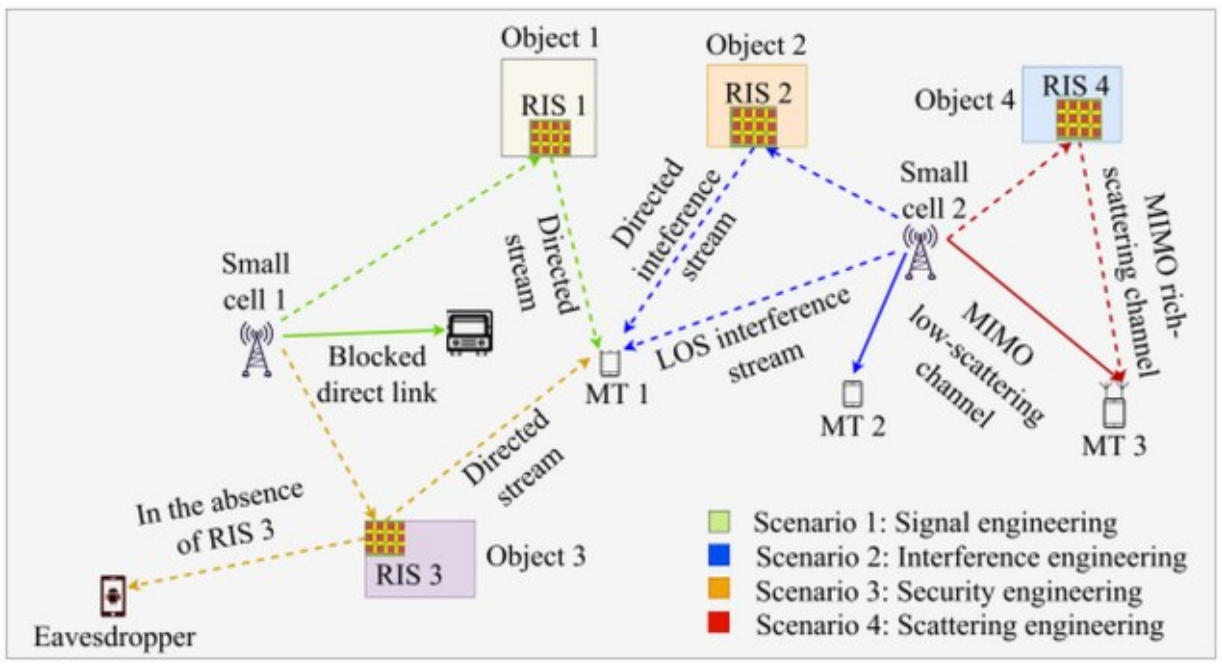

Fig. 3. Example of smart radio environment [4].

One of the most critical challenges in RIS-aided systems is scalability. Achieving meaningful performance improvements often requires large surfaces composed of hundreds or thousands of tunable elements. However, as the RIS size increases, the complexity of configuration, control signaling, and optimization grows rapidly [3]. The overhead associated with channel state information (CSI) acquisition and RIS configuration can become prohibitive, particularly in dense networks with multiple users and dynamic channels. This challenge is further exacerbated in high-frequency bands, such as millimeter-wave and terahertz communications, where channel coherence times are short and hardware imperfections are more pronounced [4], [5].

At the hardware level, non-idealities and implementation constraints pose additional challenges. Practical RIS elements exhibit finite phase resolution, insertion losses, mutual coupling, and frequency-dependent responses that deviate from the idealized models commonly assumed in theoretical studies [6], [7]. These effects can significantly degrade system performance and lead to mismatches between predicted and observed gains. Moreover, power consumption associated with RIS control circuits, especially in active or semi-active designs, raises concerns regarding energy efficiency and thermal management in large-scale deployments [8].

To address these challenges, a range of promising solutions has been proposed in recent research. From an algorithmic perspective, low-complexity and scalable optimization techniques are being developed to reduce the computational burden associated with RIS configuration. Examples include element grouping, hierarchical control architectures, and approximation-based optimization methods that trade marginal performance loss for substantial reductions in overhead [9], [10]. These approaches enable practical operation even when only partial or imperfect CSI is available.

Learning-based methods are emerging as another powerful tool for tackling complexity and adaptability challenges. Machine learning and reinforcement learning techniques have been applied to RIS configuration, channel estimation, and joint transceiver–RIS optimization, particularly in dynamic and mobile environments [11]–[13]. By leveraging historical data and online interaction with the environment, learning-based approaches can significantly reduce training overhead and enable near real-time adaptation. Hybrid strategies that combine physics-based models with data-driven learning are especially promising, as they balance interpretability, generalization, and performance robustness [11], [12].

From a hardware standpoint, advances in materials, device technologies, and circuit integration are expected to alleviate several practical constraints. The development of low-loss tunable elements, efficient biasing networks, and compact control electronics can improve RIS efficiency and reliability [6], [7]. In parallel, active and hybrid RIS architectures, which incorporate amplification or signal regeneration capabilities, are being explored to overcome path-loss limitations, albeit at the cost of increased power consumption and noise [8], [9]. Careful co-design of hardware and control algorithms is therefore essential to fully exploit these architectures.

Looking toward the future, RIS is expected to play an increasingly important role in 6G and beyond wireless networks. Integration with emerging paradigms such as integrated sensing and communication, cell-free massive MIMO, and intelligent network orchestration is likely to unlock new functionalities and performance gains [1], [2], [5]. Furthermore, RIS-enabled environments may facilitate novel services beyond traditional communications, including high-accuracy localization, environmental sensing, and physical-layer security enhancement [9], [10].

In addition, standardization efforts and realistic large-scale field trials will be crucial for transitioning RIS from experimental prototypes to deployable infrastructure. Establishing common models, performance metrics, and control interfaces can significantly accelerate industrial adoption and interoperability [3], [6]. Overall, addressing current challenges through coordinated advances in hardware design, algorithm development, and system integration will be key to realizing the full potential of RIS as a foundational technology for future smart radio environments [1], [2].

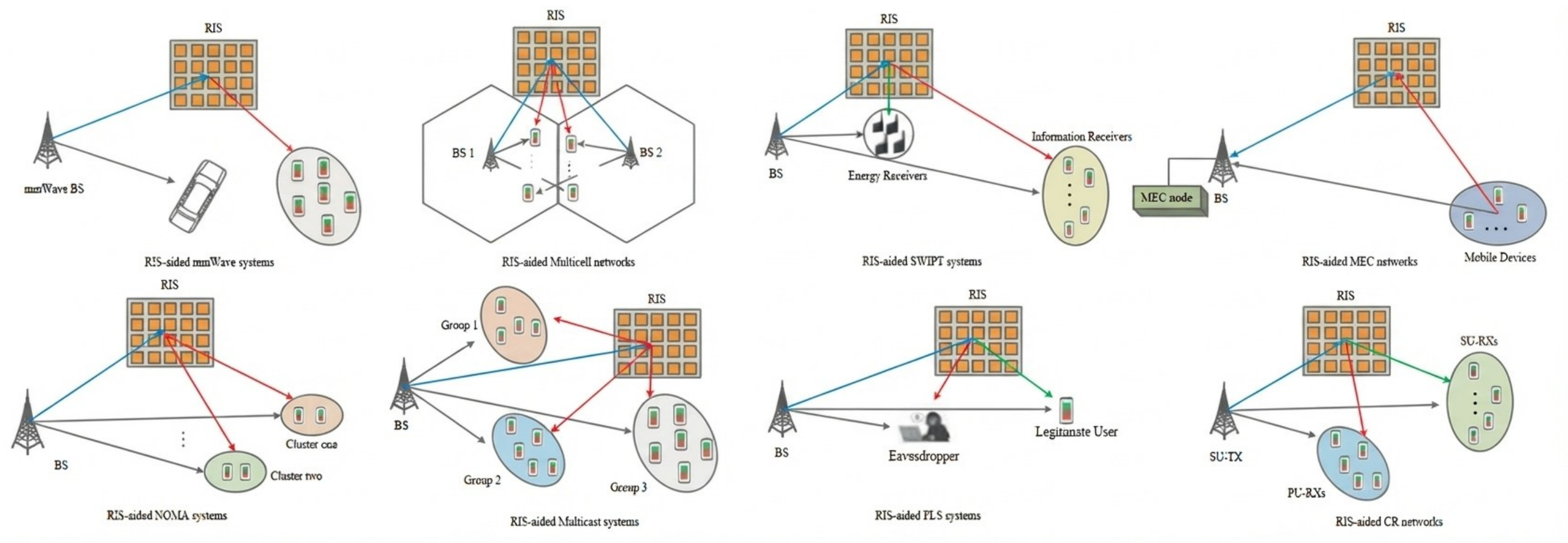

Fig. 4. Typical applications of RIS in various emerging sub-6 GHz systems [2].

## V. Conclusion

This article provided a comprehensive review of reconfigurable intelligent surfaces by bridging electromagnetic design principles with system-level wireless communication perspectives. By surveying recent advancements in RIS architectures, modeling frameworks, and emerging design methodologies, we highlighted the importance of unified abstractions that capture both physical realizations and network-level performance. Key challenges related to scalability, hardware non-idealities, channel acquisition, and control overhead were discussed, alongside promising algorithmic, hardware, and learning-assisted solutions aimed at mitigating these limitations. Looking ahead, RIS is expected to play a pivotal role in shaping 6G and beyond wireless ecosystems, particularly through its integration with intelligent network control, integrated sensing and communication, and smart radio environments. Ultimately, the transition of RIS from theoretical concepts to deployable infrastructure will depend on continued co-design across electromagnetic, circuit, and system layers, enabling practical, efficient, and scalable intelligent wireless environments.